\documentclass[11pt,a4paper]{article}
\usepackage{jheppub}
\usepackage{mathrsfs}
\newcommand{\bea}{\begin{eqnarray}}
\newcommand{\eea}{\end{eqnarray}}
\newcommand{\be}{\begin{equation}}
\newcommand{\ee}{\end{equation}}
\newcommand{\dsfrac}{\displaystyle\frac}
\newcommand{\nn}{\nonumber}

\newcommand{\dd}{\text{d}}
\newcommand{\tpsi}{\tilde{\psi}}
\newcommand{\ttheta}{\tilde{\theta}}
\newcommand{\tphi}{\tilde{\phi}}
\newcommand{\ts}{\tilde{s}}

\newcommand{\Tr}{\text{Tr}}
\newcommand{\ra}{\longrightarrow}

\newcommand{\cp}{\mathbb{CP}}
\newcommand{\sE}{\mathscr{E}}
\newcommand{\sO}{\mathscr{O}}

\newcommand{\ads}{{\rm AdS}}
\newcommand{\cft}{{\rm CFT}}
\newcommand{\cN}{{\cal N}}

\title{M-theory and Seven-Dimensional Inhomogeneous Sasaki-Einstein Manifolds}
\author[a]{Hyojoong Kim,}
\author[a,b]{Nakwoo Kim,}
\author[a]{Sunchang Kim}
\author[a]{and Jung Hun Lee}

\affiliation[a]{Department of Physics 
and Research Institute of Basic Science, \\ Kyung Hee University, \\
Hoegi-dong, Dongdaemun-gu, 
\\ Seoul, 130-701, Korea}
\affiliation[b]{Department of Physics and Astronomy,\\ 
University of British Columbia, \\ 6224 Agricultural Road, Vancouver, 
\\ British Columbia, V6T 1Z1, Canada}

\emailAdd{nkim@khu.ac.kr}

\abstract{
Seven-dimensional inhomogeneous Sasaki-Einstein manifolds
$Y^{p,k}(KE_4)$ present a challenging example of AdS/CFT correspondence. 
At present, their field theory duals for $KE_4=\mathbb{CP}^2$ 
base are proposed only within a restricted range
$3p/2\le k \le 2p$ as ${\cal N}=2$ quiver Chern-Simons-matter theories
with $SU(N)\times SU(N)\times SU(N)$ gauge group, nine bifundamental
chiral multiplets interacting through a cubic superpotential. To further elucidate
this correspondence, we use particle approximation both at classical and 
quantum level. We setup a concrete AdS/CFT mapping of conserved quantities
using geodesic motions, and turn to solutions of scalar Laplace equation 
in $Y^{p,k}$. The eigenmodes also provide an interesting subset of Kaluza-Klein 
spectrum for $D=11$ supergravity in ${\rm AdS}_4\times Y^{p,k}$, and are
dual to protected operators written in terms of matter multiplets in the dual conformal
field theory.
}


\begin{document}
\maketitle

\section{Introduction}
Thanks to the recent proposals in terms of Chern-Simons-matter theories \cite{Bagger:2007jr,Aharony:2008ug}, 
we now have a number of concrete examples for $\ads_4/\cft_3$.  On the gravity side
the internal space of M-theory is usually given as a toric Sasaki-Einstein seven-manifold, 
while on the other side of the duality we have a $D=3, \,\cN=2$ theory whose gauge
symmetry and interactions are summarised by a quiver diagram.  For the case of
Aharony-Bergman-Jafferis-Maldacena (ABJM) model \cite{Aharony:2008ug}
 M2-branes are put on 
an orbifold $\mathbb{C}^4/\mathbb{Z}_k$, where $k$ is the inverse coupling constant. 
For other orbifolds of $\mathbb{C}^4$ whose gauge dual can be derived using
D-brane intersection models, see e.g. \cite{Imamura:2008nn}.

Except for the $\cN=6$ model \cite{Aharony:2008ug}  which is in principle
amenable to exact computations in both string theory and the gauge field theory, 
most of other $\ads_4/\cft_3$ examples are less well-understood. Usually they are 
justified only by the calculation of the vacuum moduli space for the given quiver 
Chern-Simons theory, and the fact that it agrees with the toric data of the eight-dimensional 
transverse space where the M2-branes are allowed to move. Many such duality 
``examples" can be found for instance in
\cite{Martelli:2008si,Hanany:2008cd,Ueda:2008hx,Franco:2009sp,Martelli:2009ga,Benini:2009qs}.

For improvement one can  use classical membranes as a probe. 
Rotating membrane solutions in the large energy limit can provide
nontrivial quantitative predictions for long operators in the dual field theory. 
This program was initiated by the seminal paper \cite{Gubser:2002tv}, and 
shown to give a starting point for semi-classical quantization of string theory 
in $\ads_5\times S^5$ \cite{Tseytlin:2004xa}.  Nontrivial classical membrane
solutions in $\ads_4\times S^7$ which are studied in the context of
 Chern-Simons duals can be found e.g. 
\cite{Bozhilov:2007bi,Ahn:2008gd,Bozhilov:2007wn}.

More intricate backgrounds for $\ads_4/\cft_3$ are given by $d=7$
 Sasaki-Einstein (SE) 
manifolds. SE manifolds are odd-dimensional
and their metric cone provides a singular Calabi-Yau space.  There are several 
examples of seven-dimensional SE manifolds which can be constructed
as a coset. For instance the explicit metrics of so-called 
$Q^{1,1,1},M^{1,1,1},V^{5,2}$ manifolds  have been known for many years. 
Mainly as a potential model-building tool for particle physics, 
the Kaluza-Klein reduction spectra for backgrounds $\ads_4\times {\cal M}_7$,  with
for instance
${\cal M}_7=S^7,Q^{1,1,1},M^{1,1,1}$ were studied extensively in the past \cite{Pope:1984ig,Duff:1986hr}. Of course in AdS/CFT such supergravity modes
correspond to supersymmetric operators whose conformal dimensions are 
protected from quantum corrections. Anomalous dimensions of many 
non-BPS operators can be computed using classical membrane solutions moving in 
$\ads_4\times {\cal M}_7$. Membranes rotating in toric SE spaces 
$Q^{1,1,1},M^{1,1,1}$, and also in non-toric $V^{5,2}$  have been studied 
and their implications on dual CFT operaors have been reported \cite{Kim:2010ck,Lee:2010hjb,q111}. Ideally one would like to compare such 
supergravity side results with genuine
field theory computations. But the dual theories are all strongly-coupled and at present
it is very difficult to extract any quantitative data except for the spectrum of supersymmetric
operators.

Then it is logically the next step to turn to  {\it inhomogeneous}
SE manifolds. Five-dimensional SE manifolds other than $T^{1,1}=SU(2)\times SU(2)/U(1)$
are first constructed explicitly in \cite{Gauntlett:2004yd}. Dubbed $Y^{p,q}$, they are 
topologically $S^2\times S^3$ and equipped in general with a cohomogeneity-1 
metric and include $T^{1,1}$ as a special case. They are also toric and have isometry
$SU(2)\times U(1)\times U(1)$, and the dual quiver gauge theories are identified 
in \cite{Martelli:2004wu}. 
It constituted a highly nontrivial check of AdS/CFT correspondence
 that the volume of $Y^{p,q}$ match exactly  with 
the purely field-theoretical computation of central charges using $a$-maximization \cite{Intriligator:2003jj,Martelli:2005tp}.  For more works on the duality involving
$Y^{p,q}$ spaces, see e.g. \cite{Bertolini:2004xf,Benvenuti:2004dy,Berenstein:2005xa,Franco:2005zu,
Bertolini:2005di,Caceres:2010ju}.

The construction of cohomogeneity-1 SE manifolds can be generalized to arbitrary higher 
dimensions \cite{Gauntlett:2004hh}.  Given a $2n$-dimensional 
regular K\"ahler-Einstein manifold, roughly speaking one can add a squashed $S^3$ 
fibration, give a SE metric to the entire $2n+3$-dimensional space, and 
make it globally regular at the same time. 
In this paper we are interested in M-theory
backgrounds $\ads_4\times {\cal M}_7$ where ${\cal M}_7=Y^{p,k}(\cp^2)$
or $Y^{p,k}(\cp^1\times\cp^1)$. Here $p,k\in \mathbb{Z}$ determine the toric data 
and for special cases $Y^{1,1}(\cp^1\times\cp^1)=Q^{1,1,1},
Y^{2,3}(\cp^2)=M^{1,1,1}$ \cite{Martelli:2008rt}. Gauge theory duals for 
$\ads_4\times Y^{p,k}(\cp^2)$ have been proposed and their vacuum moduli space 
in the mesonic branch 
is shown to match the (metric cone of) SE space for some specific range of $p,k$ \cite{Martelli:2008si}.

In this paper we take a modest start 
in the study of conjecture for $\ads_4\times Y^{p,k}$.
We analyze some 
geodesic motions and also solve the scalar Laplace equation in $Y^{p,k}(\cp^1\times\cp^1)$ 
and  $Y^{p,k}(\cp^2)$. 
Note that for $d=5$ the geodesics and their AdS/CFT interpretation was 
given in \cite{Benvenuti:2005cz}, and 
the scalar Laplace equation in $Y^{p,q}$ was studied in 
\cite{Kihara:2005nt}, whose steps we will closely follow in Sec.4.
We will establish the mapping between the
particle solutions and CFT operators, and also elucidate their conserved charges.
The solutions of Laplace equation in $Y^{p,k}$
also provide an interesting subset of Kaluza-Klein spectrum. We present some of the 
simplest nontrivial solutions explicitly, and argue they are dual to the shortest chiral 
primary operators written purely in terms of scalar fields.

This paper is organized as follows. In Sec.2 we give a short introduction to $Y^{p,k}$,
mainly to fix the notation and provide essential information. In Sec.3 we consider particle
orbiting in SE space and establish a dictionary between supergravity description and 
the quiver Chern-Simons theory. Sec.4 is the main part where we study the Laplace 
equation and present some of the lowest lying modes explicitly. We conclude in Sec.5.

\section{Sasaki-Einstein Seven-Manifolds $Y^{p,k}$}
In this paper we are interested in the aspects of ${\rm AdS}_4/{\rm CFT}_3$ correspondence for
M-theory. The eleven-dimensional metric can be written as a direct product
of a four-dimensional anti-de Sitter space and a seven-dimensional compact manifold which
is Einstein,
\begin{equation}
 ds^2_{11} = L^2 (\frac{1}{4}ds^2_4+ ds^2_7) .
\end{equation}
Both the four and seven dimensional part (with metrics $ds^2_4$ and $ds^2_7$) 
have unit radius and satisfy 
\begin{equation}
Ric_4 = -3 g_4, \quad Ric_7 = 6 g_7 .
\end{equation}

The Einstein equation is satisfied with the inclusion of a non-vanishing four-form field
$G^{(4)} = \frac{3L^3}{8}{\rm Vol}_4$. It is well-known that when $X^7$ is 
Sasakian as well as Einstein, 
or if its metric cone provides a locally Calabi-Yau space, the overall
M-theory background is supersymmetric with eight supercharges. 
The simplest such examples
are $Q^{1,1,1}$ and $M^{1,1,1}$. These manifolds are toric, homogeneous, and 
can be considered as natural generalizations of the (base of) conifold $T^{1,1}$ to seven 
dimensions. The Kaluza-Klein reduction spectra can be found 
in ref.\cite{Pope:1984ig}. 
Their dual CFTs as supersymmetric Chern-Simons matter theory
are proposed in refs. \cite{Franco:2009sp,Martelli:2008si,Hanany:2008cd}. 
Classical solutions of rotating membranes in those
backgrounds are constructed for instance in  \cite{Kim:2010ck,q111}.
$Q^{1,1,1}$ and $M^{1,1,1}$ can be also treated as special limiting cases of the 
generically inhomogeneous $Y^{p,k}$ manifolds which are our main 
interest in this paper. For completeness let us record their metrics 
here. 
$Q^{1,1,1}$ is a twisted $U(1)$ fibration over $\cp^1 \times \cp^1 \times \cp^1$
with 
\be
ds_7^2 = \frac{1}{16} (d\psi +\sum_{i=1}^3\cos\theta_i d\phi_i)^2 + \frac{1}{8} 
\sum_{i=1}^3(d\theta_i^2 + \sin^2\theta_i d\phi_i^2) \, , 
\ee
and satisfies $R_{mn}=6g_{mn}$. The coordinates range as
 $0\le\theta_i\le\pi$, $0\le\phi_i\le2\pi$ and $0\le\psi\le4\pi$.
On the other hand
$M^{1,1,1}$ is a twisted $U(1)$ fibration over $\cp^2\times \cp^1$, with metric
\begin{align}
 ds^2_7 = &\frac{1}{64}(d\psi + 3\sin^2\mu (d\tpsi + \cos\ttheta d\tphi) + 2\cos\theta d\phi )^2 + 
 \frac{1}{8}(d\theta^2 + \sin^2\theta d\phi^2) \nn\\
            &+ \frac{3}{4}(d\mu^2 + \frac{1}{4}\sin^2\mu(d\ttheta^2 + \sin^2\ttheta d\tphi^2 + \cos^2\mu (d\tpsi + \cos\ttheta d\tphi)^2)) , 
\label{cp2}
\end{align}
where $0\le\theta,\,\ttheta\le\pi$, $0\le\phi,\,\tphi\le2\pi$, $0\le\psi,\,\tpsi\le4\pi$ and $0\le\mu\le\pi/2$.

Now let us turn to the inhomogeneous case, the so-called $Y^{p,k}$. They are 
higher dimensional generalization  
of the five-dimensional inhomogeneous
Sasaki-Einstein manifolds $Y^{p,q}$ \cite{Gauntlett:2004yd}.  
They are cohomogeneity one, 
and their geometry in arbitrary odd dimensions is studied in ref. \cite{Gauntlett:2004hh}.
We follow the formulas of ref. \cite{Gauntlett:2004hh} but specialize to 
seven dimensions. In our convention the metric is written as 
\bea
ds^2_7 = \frac{x}{4}d\ts^2_4 + 
\frac{1}{4U(x)}d x^2 + q(x)(d\psi + A)^2 + \frac{w(x)}{16}(d \alpha + f(x)(d\psi + A))^2\label{y7metric} .
\eea
The various symbols in the metric tensor are given as follows. 
\begin{align}
  U(x)&=\frac{-3x^4+4x^3+\kappa}{3x^2} ,\\
w(x)  &= U(x)+(x-1)^2 \\
&=\frac{-2x^3+3x^2+\kappa}{3x^2}, \\
    q(x)&=\frac{U(x)}{16w(x)} \\
&=\frac{-3x^4+4x^3+\kappa}{16(-2x^3+3x^2+\kappa)}, \\
  f(x)&=\frac{U(x)+x^2-x}{w(x)} \\
        &=\frac{x^3+\kappa}{-2x^3+3x^2+\kappa}.
\end{align}
One can check this metric indeed satisfies the Einstein condition $R_{mn}=6g_{mn}$ locally,
if the four dimensional manifold ${\cal M}_4$ with metric $d\ts^2_4$ is 
itself Einstein with $Ric_4=2g_4$. In fact ${\cal M}_4$ is also a K\"ahler 
manifold, and $ \tfrac{1}{2}\dd A$ should give its K\"ahler two-form.
In order to have a positive definite metric, the range of $x$ is determined by
the positivity of $U(x)$. If we define $H(x)\equiv x^4-\frac{4}{3}x^3-\frac{\kappa}{3}$, 
 $H(x)=0$ allows two (different) positive roots for $-1<\kappa<0$. 
Other two roots are complex-valued, and if we call the real roots $x_1,x_2 \,(x_1<x_2)$ they
satisfy $x_1<1<x_2$. We wish to have a smooth manifold with range
 $x_1\le x \le x_2$, by 
giving appropriate periodicity conditions to the angular coordinates $\alpha, \psi$.
This was shown to be possible in ref. \cite{Gauntlett:2004hh}, if $\kappa$ satisfies
the following conditions. The real roots $x_1,x_2$ should satisfy 
\begin{align}
3p^3x_1^3+2p^2(6b-5p)x_1^2+p(18b^2-28pb+11p^2)x_1+4(3b^3+4p^2b-6pb^2-p^3)&=0 ,\nn \\
3p^3x_2^3+2p^2(p-6b)x_2^2+p(18b^2-8pb+p^2)x_2+4b(3pb-3b^2-p^2)&=0 .\nn
\end{align}
Here $k,p$ are integers, $b=k/h$, and $h$ is the greatest common divisor of all
Chern numbers for the base ${\cal M}_4$. To be explicit we will consider 
two examples: 
$\cp^1\times\cp^1$ with $h=2$, or $\cp^2$ with $h=3$. 
$-1<\kappa<0$ is now translated to $hp/2<k<hp$.
The periodicity of various angles are given as $0\le \psi\le 2\pi$ and 
$0\le \alpha \le 2\pi l$. Regularity of the metric requires \cite{Gauntlett:2004hh}
\begin{align}
l=\frac{x_2-x_1}{p(x_2-1)(1-x_1)} ,  \quad\quad
\frac{x_1(x_2-1)}{x_2(x_1-1)}=1-\frac{hp}{k} . 
\label{regcon}
\end{align}

The form of the metric in eq.(\ref{y7metric}) is best establishing the 
regularity of $Y^{p,k}$, but it is not convenient to check the supersymmetry
or the fact it is Sasaki-Einstein. In the canonical form, the metric is locally 
written as a twisted U(1) fibration over K\"ahler-Einstein space. The 
constant norm Killing vector from the U(1) fibration is called the Reeb vector
and corresponds to the R-symmetry of the dual CFT. It can be seen through
a simple change of variables
\be
\alpha = -\phi -4\psi' ,  \quad 
  \psi = 4\psi' . 
\ee
Then the metric becomes $ds^2_7 = (d\psi'+\sigma)^2+ds^2_{KE_6}$, 
with 
\bea
ds^2_{KE_6} &=& \frac{x}{4} d\ts^2_4 + \frac{1}{4U(x)}dx^2 + \frac{U(x)}{16}(d\phi - A)^2  , \label{ke6}
\\
\sigma &=& \frac{1}{4}A + \frac{1-x}{4}(d\phi - A) . 
\eea
Note also that the Reeb vector is 
$\partial_{\psi'} = 4(\partial_{\psi} - \partial_{\alpha}) 
                 $ 
and the $KE_6$ base in eq.(\ref{ke6}) satisfies $Ric_6=8g_6$.

For definiteness and easier reference, we record here the metric and Ricci potential
for ${\cal M}_4$. When it is $\cp^1\times\cp^1$, we choose the ordinary 
spherical coordinates
\begin{align}
d\ts^2_4 &= \frac{1}{2}(d\theta_1^2 + \sin^2\theta_1 d\phi_1^2 + d\theta_2^2 + \sin^2\theta_2 d\phi_2^2) ,  \\
         A &= \cos\theta_1 d\phi_1 + \cos\theta_2 d\phi_2 . 
\end{align}
Or for $\cp^2$,  we adopt the following convention
\bea
d\ts_4 &=& 3 \biggl\{d\mu^2 + \frac{1}{4}\sin^2\mu\biggl(d\ttheta^2 + \sin^2\ttheta d\tphi^2 + \cos^2\mu(d\tpsi + \cos\ttheta d\tphi)^2\biggr)
 \biggr\} , 
\label{cp2metric}
\\
A &=& \frac{3}{2}\sin^2\mu(\cos\ttheta\dd\tphi + \dd\tpsi) . 
\eea
where $0\le\mu\le\frac{\pi}{2}$, $0\le\ttheta\le\pi$, $0\le\tphi\le 2\pi$ and $0\le\tpsi\le 4\pi$.

\section{AdS/CFT relation and Geodesic motions}
\label{3}
\begin{figure} 
\centering
\includegraphics[scale=0.28,trim= 20 40 40 40,clip=true]{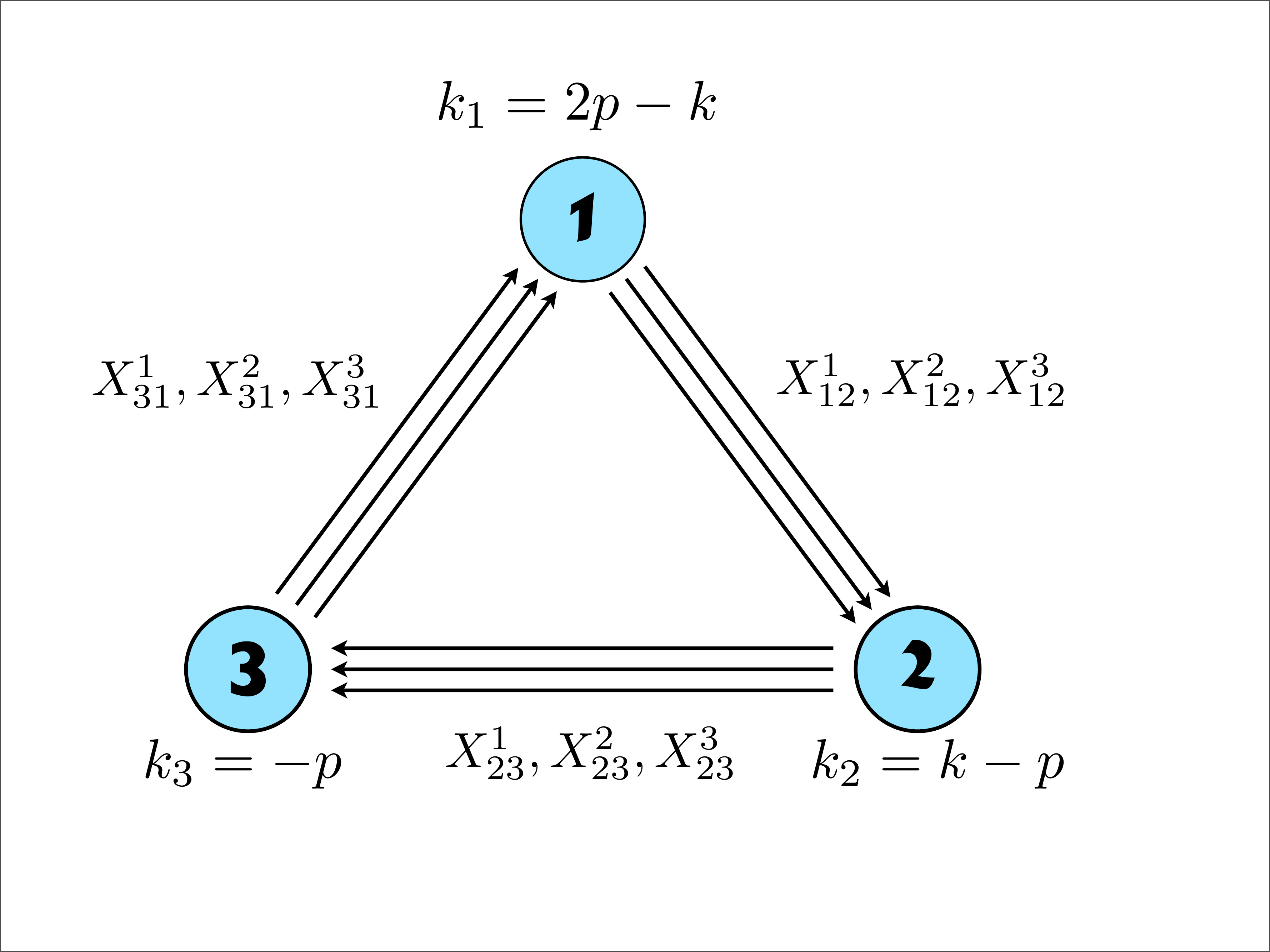}
\caption{The quiver diagram for Chern-Simons dual of $AdS_4 \times Y^{p,k}(\cp^2)$}
\label{quiver}
\end{figure}

Unlike the case of homogeneous Sasaki-Einstein seven-manifolds where 
the dual CFTs are relatively better-established 
and there exist further exploration
of the duality relation \cite{Martelli:2008rt,
Hanany:2008cd,Martelli:2008si,Kim:2010ck,q111,Lee:2010hjb}, 
the inhomogeneous examples are not very well
understood. The dual CFT of ${\rm AdS}_4\times Y^{p,k}(\cp^2)$ is
proposed in \cite{Martelli:2008si}. The field theory has gauge group
$SU(N)\times SU(N)\times SU(N)$ with Chern-Simons levels $(2p-k,k-p,-p)$. The quiver
diagram is given in figure \ref{quiver}.

There are nine chiral multiplets in total which are represented by 
arrows in the quiver diagram. They interact via a cubic superpotential
\bea
W = \sum_{i,j,k=1}^3 \epsilon_{ijk}\Tr(X_{12}^i X_{23}^j X_{31}^k).
\eea
Not surprisingly this proposal is very similar to that of $M^{1,1,1}=Y^{2,3}(\cp^2)$ which 
is a homogeneous Sasaki-Einstein seven-manifold with $KE_6=
\cp^2\times\cp^1$. It is obvious that if $p=2r,k=3r$ the quiver Chern-Simons theory 
becomes identical to the proposed dual for 
${\rm AdS}_4\times M^{1,1,1}/\mathbb{Z}_r$.
This happens when $\kappa=0$, when  $H(x)=0$ develops double roots. 
It is less clear how to see this
special arrangement leads to the homogeneous metric, $M^{1,1,1}$,
or $Q^{1,1,1}$ when ${\cal M}_4=\cp^1\times\cp^1$. 
It involves reviving 
another parameter in the metric which was originally scaled away, and taking 
a particular
 scaling limit. For details readers are referred to ref. \cite{Gauntlett:2004hh}. 

The vacuum moduli space $\mathscr{M}_3$ 
of the above ${\cal N}=2$ Chern-Simons theory
has been computed in  ref. \cite{Martelli:2008si}. When the toric data
is compared to that of $Y^{p,k}(\cp^2)$,  one finds agreement for the range 
\be
3p/2\le k \le 2p . 
\ee 
Outside this region, i.e. if $2p<k<3p$, among the toric data
of $\mathscr{M}_3$ there is one vertex which lies outside the polytope 
for $Y^{p,k}(\cp^2)$ \cite{Martelli:2008si}. To the best of our knowledge, the dual CFTs for the range of $2p<k<3p$ 
are not known yet.

Let us consider the lowest-level chiral primary operators which are  
written purely 
in terms of the scalar fields of the CFT. They constitute the lowest-lying
modes of Kaluza-Klein reduction of 11-dimensional supergravity on $Y^{p,k}$. 
As usual, the chiral primary operators are gauge singlets and classified up to F-term condition. 
The simplest ones we can think of are 
\be
(\mathscr{O}^{3}_0)^{ijk} = \Tr (X_{12}^i X_{23}^j X_{31}^k ) . 
\ee
Due to the F-term conditions the $SU(3)$ indices $i,j,k$ are symmetrized, 
so these operators are in ${\bf 10}$ of $SU(3)$. Being of the same order as $W$ and BPS, 
the conformal dimension $\Delta$ and R-charge $R$ are both 2. There is one
more global charge we can match against the geometric data, which is the monopole
charge number $Q_m$. Since we do not have any monopole operator insertion, for 
$\mathscr{O}^{3}_0$ we set $Q_m=0$.

In the reconstruction of the geometry of Sasaki-Einstein space, it is crucial 
to incorporate the monopole operators. The diagonal one $e^{ia}$, 
which is supersymmetric
and does not carry bare conformal dimension or R-charge, has charge vector
(for abelian case) 
exactly the same as the Chern-Simons levels. For the quiver theory of
figure \ref{quiver} it is $(2p-k,k-p,-p)$. It is then easily seen that 
\be
\mathscr{O}^k_+ = \Tr ( e^{ia} X^{k-p}_{12} X^p_{31} ) 
\ee
is a neutral operator. Note that we are schematic here and the symbol $\Tr$
means contracting various indices of $e^{ia},X_{12},$ and $X_{31}$ appropriately 
so that we have a gauge singlet in the end. We also have suppressed the $SU(3)$ indices
but it is understood that they are symmetrized due to F-term condition, like $\mathscr{O}^3_0$.
 Being supersymmetric,  the conformal dimension and R-charge should be still the same and we set $\Delta=R=r_+$. Here $r_+$ is not known yet but will be fixed using AdS/CFT correspondence.
The monopole
number is $Q_m=1$. 

In the same way we can think of 
\be
\mathscr{O}_-^{3p-k} = \Tr ( e^{-ia} X^{2p-k}_{12} X^p_{23} ) , 
\ee
with $\Delta=R=r_-$ and $Q_m=-1$.  At this stage we only know 
\be
r_+ + r_- = 2p . 
\label{r_id}
\ee

One can construct higher-level chiral primary operators by taking symmetric
products of the three basic operators (to be precise multiplying
 the expressions within $\Tr$, and taking 
the trace after multiplication to have a single-trace operator) given above. 
And such operators are dual to orbiting particles in the supergravity background 
of ${\rm AdS}_4\times Y^{p,k}$. More concretely, we consider particles moving only along 
the $U(1)$ fibre in the canonical form. In other words we consider geodesic
motions with ansatz $t=\kappa \tau, \, \psi'=\omega\tau$ and set all the
remaining angles including $x$ to constant. The computation is elementary 
and we obtain $\kappa=\pm 2\omega$. We restrict to holomorphic expressions 
of the complex scalar fields and choose $\kappa=2\omega>0$. 
For our purpose it is important to 
have the ratios between varioius conserved charges. We define $E$ to be the
conjugate momentum for $t$, 
$J_{\psi'}$ as conjugate to $\psi'$ etc.  
For $Y^{p,k}(\cp^2)$, 
\bea
E:J_{\tphi}:J_{\tpsi}:J_{\phi}:J_{\psi'}=1:\frac{3}{4}x\sin^2\mu\cos\ttheta:\frac{3}{4}x\sin^2\mu:\frac{1-x}{2}:2 . 
\eea
Note that $x,\mu,\ttheta$ are constants here.
The ratios can take values within a limited range, for instance 
$0\le J_{\tphi}\le \tfrac{3}{4} x_2 E$. 
 On the other hand, for $Y^{p,k}(\cp^1
\times\cp^1)$,  we obtain 
\bea
E:J_{\phi_1}:J_{\phi_2}:J_{\phi}:J_{\psi'}=1:\frac{x}{2}\cos\theta_1:\frac{x}{2}\cos\theta_2:\frac{1-x}{2}:2,
\eea
where $x,\theta_1,\theta_2$ are constants. 

We next consider matching the CFT side data for chiral primaries and
the gravity side data from geodesic motion, for $Y^{p,k}(\cp^2)$.
 On the CFT side, we have five commuting
physical observables which may have non-trivial values for operators such 
as $\mathscr{O}^k_0,\mathscr{O}_+^k,\mathscr{O}_-^{3p-k}$.
They are $\Delta, R, Q_m$ and also two more charges which determine the $SU(3)$
representation. Let us first identify $\Delta$ with $E$. 
Then from the fact that $E:J_{\psi'}=1:2$ it is obvious we should relate
$R=J_{\psi'}/2$.  Changing $\mu,\ttheta$
should correspond to assigning different $SU(3)$ indices, 
since they are among $\cp^2$ angles. We will thus identify $J_{\tilde{\phi}},J_{\tilde{\psi}}$
with the Cartan generators of the $SU(3)$ symmetry.
It turns out correct to relate $J_{\phi}$ with the monopole 
number $Q_m$. For $x=1$ orbits,  we have $Q_m=0$ and the duals are  
without monopole operator insertions. $x=x_2$ orbits are in fact 
dual to operators with maximally possible $e^{ia}$ insertions, like $\mathscr{O}^{k}_+$.
In the same way we should identify operators like $\mathscr{O}^{3p-k}_-$ with $x=x_1$
orbits.  We provide more concrete $SU(3)$ part identifications and  
check our mappings with several examples in the following.

Let us first consider $x=1$ cases and try to find out the relation between 
the highest weights of $SU(3)$ representation and angular momenta of particle solution. 
For $SU(3)$ 
we follow the standard convention and use 
\be
Q_3 = \frac{1}{2} \begin{pmatrix}1 & 0 & 0 \\ 0 & -1 & 0 \\ 0 & 0 & 0  \end{pmatrix}
, \quad
Q_8 = \frac{1}{2\sqrt{3}} \begin{pmatrix}1 & 0 & 0 \\ 0 & 1 & 0 \\ 0 & 0 & -2  \end{pmatrix} ,
\label{gel}
\ee
for fundamental representation. 
 For instance if we consider $\mathscr{O}^3_0$
with $i=j=k=1$,  we obtain $Q_3=3/2, Q_8=\sqrt{3}/2$. In general for a symmetric
product like $\Tr [( X_{12} X_{23} X_{31} )^n]$ with all $SU(3)$ indices equal to 
$1$,  we would
obtain $\Delta=R=2n, Q_3=3n/2, Q_8=\sqrt{3}n/2$. 
 Now looking at the metric
convention for $\cp^2$ in  eq.\eqref{cp2metric},
we want to 
identify these operators with orbits at $\mu=\pi/2,\ttheta=0$. In a similar way, 
$i=2$ maps to $\mu=\pi/2, \ttheta=\pi$, and $i=3$ is for $\mu=0$. When we consider 
the possible values for ratios from the gravity side and the eigenvalues of 
$Q_3,Q_8$, it is not difficult to conclude that we should identify 
\bea
Q_3 &=& J_{\tilde{\phi}} ,  \\
Q_8 &=&  \sqrt{3}(J_{\tpsi}+J_{\phi}-\frac{1}{4}J_{\psi'})  . 
\eea
This part of the consideration is very similar to the $M^{1,1,1}$ case \cite{Kim:2010ck}. 

Now what about operators with monopoles, like $\mathscr{O}_+^{k}, \mathscr{O}_-^{3p-k}$?
As mentioned earlier, 
we assume $x=x_2$ orbits have maximally possible insertions of $e^{ia}$, like
$\Tr [  ( e^{ia} X^{k-p}_{12} X^p_{31} )^n]$. And $x=x_1$ orbits are dual to operators like
$\Tr [ ( e^{-ia} X^{2p-k}_{12} X^p_{23} )^n ]$. We check this conjecture leads to 
a nontrivial realization of eq.\eqref{r_id}. One can now fix the values $r_+,r_-$ by 
 considering the ratio $J_{\tphi}/E$ and
$Q_3/\Delta$ for $\mathscr{O}_+^k,\mathscr{O}_-^{3p-k}$. 
\be
r_- = \frac{2(3p-k)}{3x_1}, \quad r_+ = \frac{2k}{3x_2} . 
\label{rpm}
\ee
One can easily see that this is consistent with eq.\eqref{r_id}, with the help
of the second identity in eq.\eqref{regcon}.  We have the same type 
of consistency when considering $Q_8/\Delta$. 
Finally we need to fix the 
proportionality coefficient in identifying $J_\phi$ with $Q_m$. It turns out we should
relate
\be Q_m = - l J_\phi ,
\ee
which implies $r_+ = \frac{2}{l(x_2-1)}, r_-=\frac{2}{l(1-x_1)}$ from the 
consideration of  $\mathscr{O}^k_+,\mathscr{O}^{3p-k}_-$. 
This assignment is easily shown to be identical to eq.\eqref{rpm}, using eq.\eqref{regcon}.

\section{Scalar Laplacian on $Y^{p,k}$}
\subsection{Scalar Laplacian and Kaluza-Klein reduction}
We now turn to the solutions of Laplace equation for $Y^{p,k}$. There are two 
motivations for doing this. One is as the quantum mechanism on $Y^{p,k}$. 
In order to explore the AdS/CFT correspondence, in principle we need to {\it quantize}
the membrane action in the nontrivial background ${\rm AdS}_4 \times Y^{p,k}$. 
This is certainly a very nontrivial problem, and one can alternatively tackle 
quantization of particle motion and try to obtain some (limited) information 
on M-theory spectrum at quantum level. 

Another motivation is as part of the Kaluza-Klein (KK) reduction problem. 
As an example of ${\rm AdS}_4/{\rm CFT}_3$ correspondence, one needs to perform the KK
computation and obtain the matter fields for four-dimensional supergravity. According
to AdS/CFT, these
supergravity modes are dual to chiral primary operators in the dual field theory. The entire
KK computation is not a trivial task, but we have complete results for $S^7$ and other 
coset manifolds such as $Q^{1,1,1},M^{1,1,1}$ \cite{Duff:1986hr,Pope:1984ig,Fabbri:1999mk}.
The space of our interest $Y^{p,k}$ is not a coset nor homogeneous, and the KK spectrum 
is not dictated by symmetry through group theory computations. Instead of the full analysis,
we will consider a simpler subset, i.e. the scalar Laplacian in this paper. 

Although the eleven-dimensional supergravity does not have any scalar field, 
the spectrum of scalar Laplacian makes an appearance in KK computation. On the problem 
of separating the various Laplace-Beltrami equations for metric tensor and four-form flux, 
readers are referred to a classic review paper on Kaluza-Klein 
supergravity by Duff et al. \cite{Duff:1986hr}. Their computation is summarised in table 5 of
\cite{Duff:1986hr}, 
and the scalar Laplacian among other things 
gives rise to the modes called $0^{+(1)}$, with 
four-dimensional mass
\be
          m^2 = \mathscr{E} + 44 - 12\sqrt{\sE + 9}  . 
\ee
Our convention is $\Box Y = -\sE Y$ with harmonic functions $Y$. 
We need to recall that in 
the convention of ref.~\cite{Duff:1986hr} 
the conformal coupling term of scalar fields 
with Ricci scalar is written separately. Then the above relation implies the 
existence of CFT operators with conformal dimension 
\be
          4\Delta(\Delta - 3) = 
\sE + 36 - 12 \sqrt{\sE+9} 
\label{cd}
, 
\ee
according to the standard AdS/CFT prescription. 

As a warm-up, let us first consider the homogeneous Sasaki-Einstein manifolds
and check eq.\eqref{cd} leads to consistent predictions on dual CFT.
\begin{enumerate}
 \item $S^7$

For round $S^7$ with unit radius, the eigenvalues are 
$\sE=j(j+6),\,\, j=0,1,2,\ldots$, for rank-$j$ totally symmetric representation 
of $SO(8)$. For dual operators we have $\Delta=j/2$. This is consistent with the 
fact that there are (allowing insertions of monopole operators) effectivly
eight scalar fields $X^I \, (I=1,2,\ldots,8)$ with $\Delta=1/2$ 
in ABJM model with Chern-Simons level $k=1$. For instance, a chiral primary operator is written as
\be
S_{I_1 I_2 \cdots I_j} \Tr (X^{I_1}X^{I_2}\cdots X^{I_j}) , \quad\quad \Delta=j/2 , 
\ee
where $S_{I_1 I_2\cdots}$ is symmetric and traceless. 
\item $Q^{1,1,1}$

The eigenvalues are computed for instance in \cite{Pope:1984ig}, 
\be 
\sE = 8(j_1(j_1+1)+j_2(j_2+1)+j_3(j_3+1)-s^2), 
\ee
where $s=0,\pm 1/2,\pm 1,\ldots$ and $j_1,j_2,j_3=|s|,|s|+1,|s|+2,\ldots$. The lowest-lying nontrivial mode
is given as $j_1=j_2=j_3=s=1/2$, or $(2,2,2)$ representation of the global symmetry 
$SU(2)\times SU(2)\times SU(2)$, with $\Delta=1$. There exist at least two 
proposals for CFT dual of (orbifolded) ${\rm AdS}_4\times Q^{1,1,1}$, see for instance  \cite{Franco:2009sp,Aganagic:2009zk}. 
And both of them exhibit 
chiral primary operators in $(2j+1,2j+1,2j+1)$ with $\Delta=2j$. The corresponding 
bulk scalar modes are identified as eigenfunction of Laplace operator with 
$j_1=j_2=j_3=s=j$. 
\item $M^{1,1,1}$

The eigenvalues are  given as \cite{Pope:1984ig}
\be
\sE =\frac{16}{3}(k^2 + 2(1+3|s|)k +6|s|)) + 8(j(j+1)-4s^2) + 64s^2, 
\ee
where $s=0,\pm\tfrac{1}{2},\pm 1, \cdots$, $j=2|s|,2|s|+1,\cdots$, and
$k=0,1,2,\cdots$. $M^{1,1,1}$ has $SU(3)\times SU(2)\times U(1)$ symmetry, and 
$j$ determines the $SU(2)$ 
representation, $s$ is for the $U(1)$ charge, and $k,s$ together determine $SU(3)$ representaion. 
In particular, for $s>0$ the eigenmodes are in $(k,k+6s)$ of $SU(3)$ and if $s<0$ the 
$SU(3)$ representation is in $(k+6|s|,k)$ \cite{Pope:1984ig}. The basic chiral primary operator 
for dual CFT is in ${\bf 10}$, or rank-3 symmetric tensor which can be also written as 
$(0,3)$-representation.  At the same time they are a triplet of $SU(2)$ and have
$\Delta=2$. This particular set of operators can be mapped to eigenmodes with 
$k=0,j=1,$ and $s=1/2$. Then we have $\sE=40$ and can match with $\Delta=2$. More generally,
if we consider symmetric products they are dual to the modes in 
$(0,6s)$-rep of $SU(3)$ and spin-$2s$ representation of $SU(2)$, for $s=\tfrac{1}{2},1,\tfrac{3}{2},
\ldots$.
\end{enumerate}

\subsection{Separation of variables and ODE with five singularities}
\subsubsection{$\cp^1\times\cp^1$ Base}
One can begin with the computation of scalar Laplace operator for the seven-manifold.
       \begin{align}
\square = &\frac{4}{x^2}\partial_x \biggl( x^2 U(x)\partial_x \biggr) + \frac{8}{x} \sum_{i=1}^2 
\biggl[ \frac{1}{\sin\theta_i}\partial_{\theta_i}(\sin\theta_i\partial_{\theta_i}) + \biggl( \frac{1}{\sin\theta_i}\partial_{\phi_i} + \cot\theta_i\partial_{\psi}\biggr)^2 \biggr] \nonumber \\
          &
 + \frac{16}{U(x)}\biggl( \partial_{\alpha} + (1-x)(\partial_{\psi} - \partial_{\alpha}) \biggr)^2 + 16(\partial_{\psi}-\partial_{\alpha})^2 . 
        \end{align}
As usual we separate the variables by writing putative eigenmodes as 
\be
\Phi(x,\theta_1,\theta_2,\phi_1,\phi_2,\psi,\alpha) =
X(x)\Theta_1(\theta_1)\Theta_2(\theta_2)\text{exp}\biggl[i\biggl(N_{\phi_1}\phi_1+N_{\phi_2}\phi_2
+N_{\psi}\psi+\dsfrac{N_{\alpha}}{l}\alpha\biggr)\biggr],
\ee
One first solves the $\cp^1$ parts one by one, using 
\begin{align}
\biggl[ \dsfrac{1}{\sin\theta_1}\partial_{\theta_1}(\sin\theta_1 \partial_{\theta_1})
 &+ \biggl( \dsfrac{N_{\phi_1}}{\sin\theta_1} + N_{\psi} \cot\theta_1 \biggr)^2
\biggr]\Theta_1
=-\biggl( j_1(j_1+1)
-N_{\psi}^2\biggr)\Theta_1
\label{cp1cp1}         
\end{align}
and also in a similar way for $\Theta_2$. The $SU(2)$ quantum numbers $j_1,j_2$ can 
take values
 $|N_{\psi}|, |N_{\psi}|+1,\cdots$.
Now the Laplace equation $\Box\Phi=-\sE\Phi$ is reduced to a
 second order ordinary differential equation (ODE) for $X(x)$, 
\bea
&&\frac{4}{x^2}\dsfrac{\dd}{\dd x}\biggl(x^2 U(x)\dsfrac{\dd}{\dd x}X(x)\biggr)-
\biggl\{\frac{8}{x} \biggl(j_1(j_1+1)+j_2(j_2+1)-2N_{\psi}^2\biggr)\nn \\
&&+\frac{16}{U(x)}\biggl(\frac{N_{\alpha}}{l}+(1-x)\biggl(N_{\psi}-\frac{N_{\alpha}}{l}\biggr)\biggr)^2 
+16\biggl(N_{\psi}-\frac{N_{\alpha}}{l}\biggr)^2-\sE\biggr\} X(x)=0.\label{odecp1cp1} 
\eea
\subsubsection{$\cp^2$ Base}
It is straightforward to compute the Laplace operator. 
      \begin{align}
\square = &\frac{4}{x^2}\partial_x \biggl( x^2 U(x)\partial_x \biggr) +
           \frac{4}{x}\biggl\{ \frac{1}{3\sin^3\mu\cos\mu}\partial_{\mu}(\sin^3\mu\cos\mu\partial_{\mu}) +
		   \frac{4}{3\sin^2\mu} \biggl[ \frac{1}{\sin\ttheta}\partial_{\ttheta}(\sin\ttheta\partial_{\ttheta}) \nonumber\\
          &\quad+ \biggl(\frac{1}{\sin\ttheta}\partial_{\tphi} - \cot\ttheta\partial_{\tpsi} \biggr)^2\biggr] +
     	  \frac{4}{3\sin^2\mu\cos^2\mu}\biggl( \partial_{\tpsi} - \frac{3}{2}\sin^2\mu\partial_{\psi} \biggr)^2 \biggr\} \nn\\
          &\quad+ \frac{16}{U(x)} \biggl(\partial_{\alpha} + (1-x)(\partial_{\psi} -\partial_{\alpha})\biggr)^2 + 
		  16(\partial_{\psi}-\partial_{\alpha})^2 .
     \end{align}
And we again employ the technique of separating the variables by 
assuming an eigenfunction of the following form.
\bea
\Phi(x,\mu,\tilde\theta,\tilde\phi,\tilde\psi,\psi,\alpha)={{X}(x){M}(\mu)\Theta(\tilde\theta)}
\text{exp}\biggl[i\biggl(N_{\tilde\phi}\tilde\phi+
\dsfrac{N_{\tilde\psi}}{2}\tilde\psi+N_{\psi}\psi+\dsfrac{N_{\alpha}}{l}\alpha\biggr)\biggr] . 
\eea

Now some of the partial derivatives turn into integration constants, and then we solve the $\cp^2$ part. 
The $\cp^1\subset \cp^2$ should be tackled first, and we can effectively substitute
\be
\frac{1}{\sin\ttheta}\partial_{\ttheta}(\sin\ttheta\partial_{\ttheta}) 
+ \biggl(\frac{1}{\sin\ttheta}\partial_{\tphi} - \cot\ttheta\partial_{\tpsi} \biggr)^2
\ra
-j(j+1)+\frac{N^2_{\tilde{\psi}}}{4} , 
\ee
where $N_{\tilde{\psi}}$ is integer, and $j=|N_{\tilde{\psi}}|/2, |N_{\tilde{\psi}}|/2+1, \cdots .$
The equation for $M(\mu)$ should complete the solution for $\cp^2$ part.  
The result is determined by the group theory for $SU(3)$, simply an eigenvalue of quadratic Casimir 
operator. We obtain
\begin{align}
\nonumber
\dsfrac{1}{3\sin^3\mu\cos\mu}\partial_{\mu}(\sin^3\mu\cos\mu\partial_{\mu})
 &- \dsfrac{ 4 j(j+1)-N_{\tilde\psi}^2}{3\sin^2\mu}
 -\frac{(N_{\tpsi} - 3 N_{\psi}\sin^2\mu )^2 }{3\sin^2\mu\cos^2\mu}
\\
&\ra-\dsfrac{4}{3}(M_1+M_2+M_1 M_2), \label{cp2lag}
\end{align}
$M_1,M_2$ determine the relevant $SU(3)$ representation. They range as 
$M_1=s$, $M_2=s+3N_{\psi}$ and $s=0,1,\cdots .$ Now we have an ordinary differential 
equation for $X(x)$,
\bea
&&\frac{4}{x^2}\dsfrac{\dd}{\dd x}\biggl(x^2 U(x)\dsfrac{\dd}{\dd x}X(x)\biggr)-
\biggl\{\frac{4}{x}\dsfrac{4}{3}\Big(s+(s+3N_{\psi})+s(s+3N_{\psi})\Big) \nn \\
&&+\frac{16}{U(x)}\biggl(\frac{N_{\alpha}}{l}+(1-x)\biggl(N_{\psi}-\frac{N_{\alpha}}{l}\biggr)\biggr)^2 
+16\biggl(N_{\psi}-\frac{N_{\alpha}}{l}\biggr)^2-\sE\biggr\} X(x)=0. \label{odecp2}
\eea

One can easily see that, not surprisingly, the ODEs eq.\eqref{odecp1cp1} and eq.\eqref{odecp2} 
are of the same form apart from the integration constants from  ${\cal M}_4$. 
Let us  introduce a new constant
\bea
Q_R &=& 2(N_\psi - \frac{N_\alpha}{l} ), 
\eea
which is the eigenvalue for $i\partial_{\psi'}/2$, so gives us the R-charge of the solution. 
On the other hand $N_\alpha$ is integral and since $il\partial_\alpha$ is related to monopole charge, 
we can interpret it as $Q_m=N_\alpha$. To simplify the ODE, we introduce a shorthand notation 
for the eigenvalues of four-dimensional Laplacian as follows
\be
C= \left\{
\begin{array}{ll}
\dsfrac{4}{3}\Big(s+(s+3N_{\psi})+s(s+3N_{\psi})\Big), & \text{for $\mathcal{M}_4=\cp^2$} . 
\cr
  2\Big( j_1(j_1+1)+j_2(j_2+1)-2N_{\psi}^2\Big),      & \text{for $\mathcal{M}_4=\cp^1\times\cp^1$} . 
\end{array}
\right.
\label{eigenvalueC}
\ee
Obvious $C$ is determined by the representation of the solution for 
non-R global symmetry. 

Then the ODE can be written as follows, 
\begin{align}
&\dsfrac{\dd^2}{\dd x^2}X(x)+\sum^4_{i=1}\frac{1}{x-x_i}\dsfrac{\dd}{\dd x}X(x)+\frac{1}{H(x)}\biggl\{-\frac{1}{9}\biggl(6\frac{N_{\alpha}}{l}-Q_R\biggr)^2  
\nn\\
&\quad\quad + x\biggl(C+\frac{2Q_R}{3}\biggl(6\frac{N_{\alpha}}{l}+Q_R\biggr)\biggr) 
 -\frac{\sE}{4}x^2-\sum^4_{i=1}\frac{\alpha_i^2H'(x_i)}{x-x_i}\biggr\}X(x)=0 . 
\end{align}
As defined earlier $H(x)= x^4-\frac{4}{3}x^3-\frac{\kappa}{3}=\prod_{i=1}^4(x-x_i)$. 
Among the roots $x_1,x_2$ are real but $x_3,x_4$ are complex-valued.
This ODE has five regular singular points on complex plane, at $x=x_1,x_2,x_3,x_4,$ and $\infty$.
The parameters $\alpha_i$ are given as for instance
\be
\alpha_1 = -\frac{\Big(4\dsfrac{N_{\alpha}}{l}+2(1-x_1)Q_R\Big)x_1^2}{2(x_1-x_2)(x_1-x_3)(x_1-x_4)}
= \frac{Q_R}{4}-\frac{N_\alpha}{2l}\frac{1}{x_1-1} , 
\ee
and similarly for others. Note that $\alpha_3,\alpha_4$ are complex-valued but they are complex
conjugate to each other. One can easily show 
\be
\sum \alpha_i = Q_R .
\ee

The asymptotic behavior of $X(x)$ near $x=x_i$ is given as $X(x)\sim (x-x_i)^{\alpha_i}$. 
If we extract the asymptotic behavior by setting 
\be
X(x) = \prod^4_{i=1}(x-x_i)^{\alpha_i}f(x),
\ee
we have the following ODE in standard form 
\be 
\frac{\dd^2}{\dd x^2}f(x)+\sum^4_{i=1}\frac{1+2\alpha_i}{x-x_i}\frac{\dd}{\dd x}f(x)
 +\frac{\alpha x^2+\beta x}{\prod_{i=1}^{4} (x-x_i)}f(x)=0 . 
\label{sform}
\ee
The parameters $\alpha,\beta$ are given as 
\begin{align}
\alpha 
       &= Q_R(Q_R+3)-\frac{\sE}{4}  , \\
 \beta &= C-4\frac{N_{\alpha}}{l}-2Q_R . 
\end{align}

\subsection{Explicit solutions and BPS conditions}
In this section we will present simple solutions for $f(x)$ which are either constant
or linear in $x$.  We also try to give their interpretation as operators in quiver 
gauge theory figure \ref{quiver}, for ${\cal M}_4=\cp^2$.

\subsubsection{Constant solutions and Chiral Primaries}
Obviously 
$f(x)=const.$ becomes a 
solution  if $\alpha=\beta=0$. $\alpha=0$  implies
$\Delta=Q_R$, which is the familiar supersymmetry condition for chiral primaries that
conformal dimension should be equal to R-charge. 
Then $\beta=0$ leads to $C=4N_\psi$. Since $N_\psi=Q_R/2+N_\alpha/l$, 
this condition relates the representation of non-R flavor symmetry with monopole
number and R-charge. 

We can easily check that 
these conditions indeed account for the chiral primary operators of $Y^{p,k}(\cp^2)$, as follows.
Let us start with the case $N_\psi=1$. Then $\beta=0$, 
or equivalently $C=4$ implies we should set $s=0$, 
i.e. the $SU(3)$ representation is in $(0,3)$, i.e. ${\bf 10}$.  If we further set 
$N_\alpha=0$ then $Q_R=2$. Now we look at $\alpha=0$ and see $\Delta=2$. This particular
state obviously corresponds to $\sO^3_0=\Tr (X_{12}X_{23}X_{31})$. We can also 
find duals for other operators. For $\sO_+^k$, we need to choose $N_\psi=k/3, N_\alpha=1$
and $Q_R=r_+$. Or for $\sO_-^{3p-k}$, one finds $N_\psi=(3p-k)/3,N_\alpha=-1,
Q_R=r_-$ do 
the job. 
 For doing this, we can make use of the following identities which can be
derived from eq.\eqref{regcon}.
\be
l = \frac{hx_2}{k(x_2-1)} = \frac{hx_1}{(k-hp)(x_1-1)} . 
\ee
Similarly we can describe all higher composite operators, which are purely made of
scalar operators with numerous insertions of monopole operators $e^{\pm ia}$.

Now it should be clear that we {\it can} find states dual to the chiral primary operators such as 
listed in Sec.\ref{3}, but how do we know that other assignments of quantum numbers are prohibited?
For instance, what would happen if we considered $N_\psi=1,N_\alpha=1$ instead of $N_\psi=1,N_\alpha=0$
which gives us $\sO^3_0$? The correct quantization is given by regularity of wavefunction, of
course. And for that matter, in practice we need to consider two things here. 
 One is the correct periodicity for various 
angles in the metric, especially $\alpha,\psi$.  The other is the convergence of the eigenmode 
at north and south pole of squashed $S^2$, i.e. $x=x_1,x_2$.

 A systematic way of determining the correct
periodicity condition, or single-valuedness of the wavefunction, is to use toric geometry. 
$Y^{p,k}$ spaces are toric when we choose the four-dimensional K\"ahler-Einstein base 
${\cal M}_4$ as toric. It is certainly the case for ${\cal M}_4=\cp^2$ or $\cp^1\times \cp^1$. 
More precisely, the metric cone of $Y^{p,k}$ is toric, in other words it is a complex four-dimensional
space and can be expressed as $U(1)^4$ fibration over a convex rational polyhedral cone. 
In order to compute the toric data it is crucial to establish a basis for an effectively 
acting torus action $\mathbb{T}^4$. The result is reported in ref.~\cite{Martelli:2008rt}, and 
for our purpose readers are asked to bring their attention to the Killing vectors $e_3,e_4$ 
in eq.(3.7) of \cite{Martelli:2008rt}. In our notation they are 
\be
e_3 = \frac{\partial}{\partial\psi} - \frac{kl}{3} \frac{\partial}{\partial\alpha} , \quad e_4 = l \frac{\partial}{\partial\alpha} . 
\ee
The fact that they are effectively acting means that their eigenvalues should be $i$ times 
an integer. In particular, it implies $N_\alpha=Q_m$ should be an integer, and so should
$N_\psi-k N_\alpha/3$.
We should also check if $X(x)=f(x)\prod(x-x_i)^\alpha$ stays finite at $x=x_1,x_2$. For constant $f(x)$, 
we should simply avoid the cases with negative $\alpha_1,\alpha_2$.

Let us use a specific example of $(p,k)=(4,7)$ here, in order
to illustrate that the above conditions really pin down the spectrum BPS operators. 
One can easily compute $\alpha_1=\tfrac{1}{2}(N_\psi+\frac{5}{3}N_\alpha)$ 
and $\alpha_2=\frac{1}{2}(N_\psi-\tfrac{7}{3}N_\alpha)$ .
Since $N_\alpha$ is integral, 
  we start with $N_\alpha=0$. Then $N_\psi$
should be non-negative to guarantee $\alpha_1,\alpha_2 \ge 0$.  These states are 
dual to $\Tr [(X_{12}X_{23}X_{31})^{N_\psi}]$. Let us now consider $N_\alpha=1$. 
Then from the consideration of $e_3$ we see 
$N_\psi \in\mathbb{Z}+ \tfrac{7}{3} $. And since we want $\alpha_2\ge 0$, we can 
only have $N_\psi = \tfrac{7}{3}, \tfrac{10}{3}, \tfrac{13}{3}, \cdots$.  One can easily
compute $C,Q_R$ for these solutions, and convince oneself that they are dual to 
$\Tr [(e^{ia} X_{12}^{2p-k}X_{23}^p) (X_{12}X_{23}X_{31})^{N_\psi-7/3} ] $. 
A similar argument holds for $N_\alpha=-1$ etc. 

We can do something similar with ${\cal M}_4=\cp^1\times\cp^1$, although we do 
not know the dual Chern-Simons theory yet. From toric data we need integrality of 
eigenvalues for the
following Killing vectors (see eq.(3.25) of ref.~\cite{Martelli:2008rt}),
\be
e_3 = \frac{\partial}{\partial\psi} - \frac{kl}{2} \frac{\partial}{\partial\alpha} , \quad e_4 = l \frac{\partial}{\partial\alpha} .  
\ee
And we also require $\alpha_1=\tfrac{1}{2}(N_\psi+\tfrac{2p-k}{2}N_\alpha),\,
\alpha_2=\tfrac{1}{2}(N_\psi-\tfrac{k}{2}N_\alpha)$ be non-negative.
Now let us consider the condition $\beta=0$. 
We immediately see that the simplest way of satisfying $C=4N_\psi$ 
is $j_1=j_2=N_\psi$. If we again start with
$N_\alpha=0$, $N_\psi$ should be an integer and we may conjecture there should be 
BPS operators $\sO^n_0$ with $j_1=j_2=N_\psi=n$ and $Q_R=2n$. Next we consider $N_\alpha=1$, and from $\alpha_2\ge 0$ we conjecture there are operators
$\sO_+^{k/2+n} \,(n=0,1,2,\cdots)$ with $j_1=j_2=N_\psi=\tfrac{k}{2}+n$ and 
$Q_R=k-2/l+2n$.  
And for $N_\alpha=-1$, in a similar way we obtain
 $j_1=j_2=N_\psi=p-k/2+n$ with $n=0,1,2,\cdots$ from 
$\alpha_1\ge 0$. R-charge is given as $Q_R = 2p-k+2n+2/l$, and we can call these
states $\mathscr{O}^{p-k/2+n}_-$.
One can certainly continue with other values of $N_\alpha$.

\subsubsection{First excited states: linear $f(x)$}
Although it seems too difficult to find a complete set of solutions to eq.~\eqref{sform}, 
it turns out we can find some excited states where $f(x)$ is a linear function. Let us try
$f(x)=x+a$, upon which the ODE becomes
\be
\sum \frac{1+2\alpha_i}{x-x_i} + \frac{\alpha x^2+\beta x}{H(x)} (x+a) = 0 . 
\label{lin}
\ee
One may make use of the following identities,
\bea
H(x) \sum \frac{1}{x-x_i} &=& H'(x) = 4x^2(x-1) ,  \\
H(x) \sum \frac{\alpha_i}{x-x_i} &=& x^2\left( Q_R x - Q_R - \frac{2N_\alpha}{l} \right) . 
\eea
Then it is a simple matter to solve eq.~\eqref{lin}. 
One first needs to set $\alpha=-2Q_R-4$,
which implies
$\Delta=Q_R+1$. For $\beta$, there are two possibilities. $\beta=0$ which gives
$C=2Q_R+4N_\alpha/l$, or $\beta=2(Q_R+2N_\alpha/l)+4$ which means
$C=4Q_R+8N_\alpha/l+4$.

Having a simple definite relation between $\Delta$ and $Q_R$, 
we expect the duals are also BPS, and even in the same supermultiplet as constant 
solutions.  Candidate operators can be made with insertions of
 fermion bilinears, which have a different 
ratio of conformal dimension and R-charge than scalar fields. 
We can again consider different values of monopole number $N_\alpha$ and check if 
the wavefunction is single-valued and finite-valued for different representations
of $SU(3)$ or $SU(2)\times SU(2)$, but we do not go into further details here. 
\section{Discussions}
In this paper we have studied the AdS/CFT duality relation for M-theory
background AdS$_4 \times {\cal M}_7$, where ${\cal M}_7$ is an inhomogeneous Sasaki-Einstein 
manifold. For concreteness we have chosen cohomogeneity-1 examples, $Y^{p,k}(\cp^2)$
and $Y^{p,k}(\cp^1\times\cp^1)$. Using simple geodesic motions we have established 
a precise mapping between supergravity and field theory, and through scalar Laplace
equation we have seen how chiral primary operators are realized as wavefunctions 
of quantum mechanics.

The  issues covered in this paper are admittedly rather limited. First of all, we have not 
tried a full treatise of Kaluza-Klein reduction involving the metric, four-form and gravitino
fields. We have only studied scalar Laplacian and certainly it is very desirable to extend to the 
entire action. Even for scalar Laplacian, we managed to obtain only some of the lowest-lying
modes. In fact one can check if there are higher-order polynomial solutions for $f(x)$
to eq.\eqref{sform}, but when one tries a quadratic polynomial for $f(x)$ it is easy to see
that it leads to inconsistency and there is no such solution. Due to supersymmetry, 
we expect there should exist higher modes with $\Delta=Q_R+2, Q_R+3, \cdots$, 
and it will be very interesting to construct such solutions explicitly.

$Y^{p,k}$ manifolds including homogeneous ones as special cases 
certainly do not exhaust all explicit Sasaki-Einstein 7-manifolds known to us. 
There exist higher-cohomogeneity examples such as $L^{p,q,r_1,r_2}$
in seven dimensions, constructed in \cite{Cvetic:2005ft,Cvetic:2005vk}. 
Back to five-dimensions, the gauge duals for $\ads_5\times L^{p,q,r}$ 
were identified in
 refs.~\cite{Franco:2005sm,Butti:2005sw}, the geodesic motions were studied
in \cite{Benvenuti:2005ja}, while the scalar
Laplace equation was studied 
in \cite{Oota:2005mr}. We hope to be able to analyze the toric geometry, 
and the Chern-Simons
duals of $L^{p,q,r_1,r_2}$ manifolds and compare the membrane
dynamics against the CFT spectra.

\acknowledgments
This work was supported by the National Research Foundation of Korea Grant funded
by the Korean Government [NRF-2009-351-C00110].
N. Kim, S. Kim and J.H. Lee are partly supported by 
National Research Foundation of Korea with grant 
No. 2010-0023121, No. 2009-0085995, 
and also through the Center for Quantum Spactime (CQUeST) of Sogang University 
with grant No. R11-2005-021. 
H. Kim, N. Kim, and J.H. Lee are
 very grateful to PITP, UBC for hospitality. 
\bibliography{ref}{}

\providecommand{\href}[2]{#2}\begingroup\raggedright\begin{thebibliography}{10}

\bibitem{Bagger:2007jr}
J.~Bagger and N.~Lambert, {\it {Gauge Symmetry and Supersymmetry of Multiple
  M2-Branes}},  {\em Phys. Rev.} {\bf D77} (2008) 065008,
  [\href{http://xxx.lanl.gov/abs/0711.0955}{{\tt arXiv:0711.0955}}].

\bibitem{Aharony:2008ug}
O.~Aharony, O.~Bergman, D.~L. Jafferis, and J.~Maldacena, {\it {$N=6$
  superconformal Chern-Simons-matter theories, M2-branes and their gravity
  duals}},  {\em JHEP} {\bf 10} (2008) 091,
  [\href{http://xxx.lanl.gov/abs/0806.1218}{{\tt arXiv:0806.1218}}].

\bibitem{Imamura:2008nn}
Y.~Imamura and K.~Kimura, {\it {On the moduli space of elliptic
  Maxwell-Chern-Simons theories}},  {\em Prog. Theor. Phys.} {\bf 120} (2008)
  509--523, [\href{http://xxx.lanl.gov/abs/0806.3727}{{\tt arXiv:0806.3727}}].

\bibitem{Martelli:2008si}
D.~Martelli and J.~Sparks, {\it {Moduli spaces of Chern-Simons quiver gauge
  theories and $AdS_4/CFT_3$}},  {\em Phys. Rev.} {\bf D78} (2008) 126005,
  [\href{http://xxx.lanl.gov/abs/0808.0912}{{\tt arXiv:0808.0912}}].

\bibitem{Hanany:2008cd}
A.~Hanany and A.~Zaffaroni, {\it {Tilings, Chern-Simons Theories and M2
  Branes}},  {\em JHEP} {\bf 10} (2008) 111,
  [\href{http://xxx.lanl.gov/abs/0808.1244}{{\tt arXiv:0808.1244}}].

\bibitem{Ueda:2008hx}
K.~Ueda and M.~Yamazaki, {\it {Toric Calabi-Yau four-folds dual to
  Chern-Simons-matter theories}},  {\em JHEP} {\bf 12} (2008) 045,
  [\href{http://xxx.lanl.gov/abs/0808.3768}{{\tt arXiv:0808.3768}}].

\bibitem{Franco:2009sp}
S.~Franco, I.~R. Klebanov, and D.~Rodriguez-Gomez, {\it {M2-branes on Orbifolds
  of the Cone over $Q^{1,1,1}$}},  {\em JHEP} {\bf 08} (2009) 033,
  [\href{http://xxx.lanl.gov/abs/0903.3231}{{\tt arXiv:0903.3231}}].

\bibitem{Martelli:2009ga}
D.~Martelli and J.~Sparks, {\it {$AdS_4/CFT_3$ duals from M2-branes at
  hypersurface singularities and their deformations}},  {\em JHEP} {\bf 12}
  (2009) 017, [\href{http://xxx.lanl.gov/abs/0909.2036}{{\tt
  arXiv:0909.2036}}].

\bibitem{Benini:2009qs}
F.~Benini, C.~Closset, and S.~Cremonesi, {\it {Chiral flavors and M2-branes at
  toric CY4 singularities}},  {\em JHEP} {\bf 02} (2010) 036,
  [\href{http://xxx.lanl.gov/abs/0911.4127}{{\tt arXiv:0911.4127}}].

\bibitem{Gubser:2002tv}
S.~S. Gubser, I.~R. Klebanov, and A.~M. Polyakov, {\it {A semi-classical limit
  of the gauge/string correspondence}},  {\em Nucl. Phys.} {\bf B636} (2002)
  99--114, [\href{http://xxx.lanl.gov/abs/hep-th/0204051}{{\tt
  hep-th/0204051}}].

\bibitem{Tseytlin:2004xa}
A.~A. Tseytlin, {\it {Semiclassical strings and AdS/CFT}},
  \href{http://xxx.lanl.gov/abs/hep-th/0409296}{{\tt hep-th/0409296}}.

\bibitem{Bozhilov:2007bi}
P.~Bozhilov, {\it {Integrable systems from membranes on $AdS_4 \times S^7$}},
  {\em Fortsch. Phys.} {\bf 56} (2008) 373--379,
  [\href{http://xxx.lanl.gov/abs/0711.1524}{{\tt arXiv:0711.1524}}].

\bibitem{Ahn:2008gd}
C.~Ahn and P.~Bozhilov, {\it {Finite-size effects of Membranes on $AdS_4\times
  S_7$}},  {\em JHEP} {\bf 08} (2008) 054,
  [\href{http://xxx.lanl.gov/abs/0807.0566}{{\tt arXiv:0807.0566}}].

\bibitem{Bozhilov:2007wn}
P.~Bozhilov and R.~C. Rashkov, {\it {On the multi-spin magnon and spike
  solutions from membranes}},  {\em Nucl. Phys.} {\bf B794} (2008) 429--441,
  [\href{http://xxx.lanl.gov/abs/0708.0325}{{\tt arXiv:0708.0325}}].

\bibitem{Pope:1984ig}
C.~N. Pope, {\it {Harmonic Expansions on Solutions of $d = 11$ Supergravity
  with $SU(3)\times SU(2) \times U(1)$ or $SU(2) \times SU(2) \times SU(2)
  \times U(1)$ Symmetry}},  {\em Class. Quant. Grav.} {\bf 1} (1984) L91.

\bibitem{Duff:1986hr}
M.~J. Duff, B.~E.~W. Nilsson, and C.~N. Pope, {\it {Kaluza-Klein
  Supergravity}},  {\em Phys. Rept.} {\bf 130} (1986) 1--142.

\bibitem{Kim:2010ck}
J.~Kim, N.~Kim, and J.~H. Lee, {\it {Rotating Membranes in $AdS_4\times
  M^{1,1,1}$}},  {\em JHEP} {\bf 03} (2010) 122,
  [\href{http://xxx.lanl.gov/abs/1001.2902}{{\tt arXiv:1001.2902}}].

\bibitem{Lee:2010hjb}
J.~H. Lee, S.~Kim, J.~Kim, and N.~Kim, {\it {Probing Non-Toric Geometry with
  Rotating Membranes}},  {\em Nucl. Phys.} {\bf B838} (2010) 238--252,
  [\href{http://xxx.lanl.gov/abs/1003.6111}{{\tt arXiv:1003.6111}}].

\bibitem{q111}
N.~Kim and J.~H. Lee, {\it {Multi-Spin Membrane Solutions in $AdS_4\times
  Q^{1,1,1}$}},  {\em Talk given at Korean Physical Society meeting, CECO,
  Changwon, 21-23 Oct 2009}.

\bibitem{Gauntlett:2004yd}
J.~P. Gauntlett, D.~Martelli, J.~Sparks, and D.~Waldram, {\it {Sasaki-Einstein
  metrics on $S^2\times S^3$}},  {\em Adv. Theor. Math. Phys.} {\bf 8} (2004)
  711--734, [\href{http://xxx.lanl.gov/abs/hep-th/0403002}{{\tt
  hep-th/0403002}}].

\bibitem{Martelli:2004wu}
D.~Martelli and J.~Sparks, {\it {Toric geometry, Sasaki-Einstein manifolds and
  a new infinite class of AdS/CFT duals}},  {\em Commun. Math. Phys.} {\bf 262}
  (2006) 51--89, [\href{http://xxx.lanl.gov/abs/hep-th/0411238}{{\tt
  hep-th/0411238}}].

\bibitem{Intriligator:2003jj}
K.~A. Intriligator and B.~Wecht, {\it {The exact superconformal R-symmetry
  maximizes a}},  {\em Nucl. Phys.} {\bf B667} (2003) 183--200,
  [\href{http://xxx.lanl.gov/abs/hep-th/0304128}{{\tt hep-th/0304128}}].

\bibitem{Martelli:2005tp}
D.~Martelli, J.~Sparks, and S.-T. Yau, {\it {The geometric dual of
  a-maximisation for toric Sasaki- Einstein manifolds}},  {\em Commun. Math.
  Phys.} {\bf 268} (2006) 39--65,
  [\href{http://xxx.lanl.gov/abs/hep-th/0503183}{{\tt hep-th/0503183}}].

\bibitem{Bertolini:2004xf}
M.~Bertolini, F.~Bigazzi, and A.~L. Cotrone, {\it {New checks and subtleties
  for AdS/CFT and a- maximization}},  {\em JHEP} {\bf 12} (2004) 024,
  [\href{http://xxx.lanl.gov/abs/hep-th/0411249}{{\tt hep-th/0411249}}].

\bibitem{Benvenuti:2004dy}
S.~Benvenuti, S.~Franco, A.~Hanany, D.~Martelli, and J.~Sparks, {\it {An
  infinite family of superconformal quiver gauge theories with Sasaki-Einstein
  duals}},  {\em JHEP} {\bf 06} (2005) 064,
  [\href{http://xxx.lanl.gov/abs/hep-th/0411264}{{\tt hep-th/0411264}}].

\bibitem{Berenstein:2005xa}
D.~Berenstein, C.~P. Herzog, P.~Ouyang, and S.~Pinansky, {\it {Supersymmetry
  Breaking from a Calabi-Yau Singularity}},  {\em JHEP} {\bf 09} (2005) 084,
  [\href{http://xxx.lanl.gov/abs/hep-th/0505029}{{\tt hep-th/0505029}}].

\bibitem{Franco:2005zu}
S.~Franco, A.~Hanany, F.~Saad, and A.~M. Uranga, {\it {Fractional Branes and
  Dynamical Supersymmetry Breaking}},  {\em JHEP} {\bf 01} (2006) 011,
  [\href{http://xxx.lanl.gov/abs/hep-th/0505040}{{\tt hep-th/0505040}}].

\bibitem{Bertolini:2005di}
M.~Bertolini, F.~Bigazzi, and A.~L. Cotrone, {\it {Supersymmetry breaking at
  the end of a cascade of Seiberg dualities}},  {\em Phys. Rev.} {\bf D72}
  (2005) 061902, [\href{http://xxx.lanl.gov/abs/hep-th/0505055}{{\tt
  hep-th/0505055}}].

\bibitem{Caceres:2010ju}
E.~Caceres, M.~N. Mahato, L.~A. Pando~Zayas, and V.~J.~G. Rodgers, {\it {Toward
  NS5 Branes on the Resolved Cone over $Y^{p,q}$}},
  \href{http://xxx.lanl.gov/abs/1007.3719}{{\tt arXiv:1007.3719}}.

\bibitem{Gauntlett:2004hh}
J.~P. Gauntlett, D.~Martelli, J.~F. Sparks, and D.~Waldram, {\it {A new
  infinite class of Sasaki-Einstein manifolds}},  {\em Adv. Theor. Math. Phys.}
  {\bf 8} (2006) 987--1000, [\href{http://xxx.lanl.gov/abs/hep-th/0403038}{{\tt
  hep-th/0403038}}].

\bibitem{Martelli:2008rt}
D.~Martelli and J.~Sparks, {\it {Notes on toric Sasaki-Einstein seven-manifolds
  and $AdS_4/CFT_3$}},  {\em JHEP} {\bf 11} (2008) 016,
  [\href{http://xxx.lanl.gov/abs/0808.0904}{{\tt arXiv:0808.0904}}].

\bibitem{Benvenuti:2005cz}
S.~Benvenuti and M.~Kruczenski, {\it {Semiclassical strings in Sasaki-Einstein
  manifolds and long operators in $N = 1$ gauge theories}},  {\em JHEP} {\bf
  10} (2006) 051, [\href{http://xxx.lanl.gov/abs/hep-th/0505046}{{\tt
  hep-th/0505046}}].

\bibitem{Kihara:2005nt}
H.~Kihara, M.~Sakaguchi, and Y.~Yasui, {\it {Scalar Laplacian on
  Sasaki-Einstein manifolds $Y^{p,q}$}},  {\em Phys. Lett.} {\bf B621} (2005)
  288--294, [\href{http://xxx.lanl.gov/abs/hep-th/0505259}{{\tt
  hep-th/0505259}}].

\bibitem{Fabbri:1999mk}
D.~Fabbri, P.~Fre, L.~Gualtieri, and P.~Termonia, {\it {M-theory on $AdS_4
  \times M^{111}$: The complete $Osp(2|4) \times SU(3) \times SU(2)$ spectrum
  from harmonic analysis}},  {\em Nucl. Phys.} {\bf B560} (1999) 617--682,
  [\href{http://xxx.lanl.gov/abs/hep-th/9903036}{{\tt hep-th/9903036}}].

\bibitem{Aganagic:2009zk}
M.~Aganagic, {\it {A Stringy Origin of M2 Brane Chern-Simons Theories}},  {\em
  Nucl. Phys.} {\bf B835} (2010) 1--28,
  [\href{http://xxx.lanl.gov/abs/0905.3415}{{\tt arXiv:0905.3415}}].

\bibitem{Cvetic:2005ft}
M.~Cvetic, H.~Lu, D.~N. Page, and C.~N. Pope, {\it {New Einstein-Sasaki spaces
  in five and higher dimensions}},  {\em Phys. Rev. Lett.} {\bf 95} (2005)
  071101, [\href{http://xxx.lanl.gov/abs/hep-th/0504225}{{\tt
  hep-th/0504225}}].

\bibitem{Cvetic:2005vk}
M.~Cvetic, H.~Lu, D.~N. Page, and C.~N. Pope, {\it {New Einstein-Sasaki and
  Einstein spaces from Kerr-de Sitter}},  {\em JHEP} {\bf 07} (2009) 082,
  [\href{http://xxx.lanl.gov/abs/hep-th/0505223}{{\tt hep-th/0505223}}].

\bibitem{Franco:2005sm}
S.~Franco {\em et.~al.}, {\it {Gauge theories from toric geometry and brane
  tilings}},  {\em JHEP} {\bf 01} (2006) 128,
  [\href{http://xxx.lanl.gov/abs/hep-th/0505211}{{\tt hep-th/0505211}}].

\bibitem{Butti:2005sw}
A.~Butti, D.~Forcella, and A.~Zaffaroni, {\it {The dual superconformal theory
  for L(p,q,r) manifolds}},  {\em JHEP} {\bf 09} (2005) 018,
  [\href{http://xxx.lanl.gov/abs/hep-th/0505220}{{\tt hep-th/0505220}}].

\bibitem{Benvenuti:2005ja}
S.~Benvenuti and M.~Kruczenski, {\it {From Sasaki-Einstein spaces to quivers
  via BPS geodesics: $L^{pqr}$}},  {\em JHEP} {\bf 04} (2006) 033,
  [\href{http://xxx.lanl.gov/abs/hep-th/0505206}{{\tt hep-th/0505206}}].

\bibitem{Oota:2005mr}
T.~Oota and Y.~Yasui, {\it {Toric Sasaki-Einstein manifolds and Heun
  equations}},  {\em Nucl. Phys.} {\bf B742} (2006) 275--294,
  [\href{http://xxx.lanl.gov/abs/hep-th/0512124}{{\tt hep-th/0512124}}].

\end{thebibliography}\endgroup
\end{document}